\documentclass[table]{article}

\usepackage[utf8]{inputenc}
\usepackage{authblk}
\usepackage{arxiv}
\usepackage[colorlinks=true,linkcolor=blue]{hyperref}

\usepackage{graphicx}
\usepackage{tikz}
\usetikzlibrary{automata, arrows}
\usetikzlibrary{decorations.pathreplacing,angles,quotes} 
\usetikzlibrary{positioning}
\usetikzlibrary{shapes,arrows}
\usetikzlibrary{decorations.shapes}
\usetikzlibrary{arrows.meta}
\usetikzlibrary{backgrounds, fit}
\usetikzlibrary{calc}
\tikzstyle{decision} = [diamond, draw, fill=blue!20, 
text width=4.5em, text badly centered, node distance=3cm, inner sep=0pt]
\tikzstyle{block} = [rectangle, draw, fill=blue!20, 
text width=5em, text centered, rounded corners, minimum height=4em]
\tikzstyle{line} = [draw, -latex']
\tikzstyle{cloud} = [draw, ellipse,fill=red!20, node distance=3cm,
minimum height=2em]

\usepackage{graphicx}
\usepackage{subcaption}
\usepackage{enumitem}
\usepackage{algpseudocode}
\usepackage{algorithm}
\usepackage{amssymb}
\usepackage{mathtools}
\usepackage{xcolor}
\usepackage{bbold}

\begin{document}

\title{Entanglement in a 20-Qubit Superconducting Quantum Computer 
}

\author[1]{Gary J. Mooney}
\author[1]{Charles D. Hill}
\author[1, 2]{Lloyd C.L. Hollenberg}


\affil[1]{School of Physics \protect\\
          University of Melbourne \protect\\
          Parkville 3020, AUSTRALIA}
\affil[2]{lloydch@unimelb.edu.au}

\date{Received: date / Accepted: date}

\maketitle

\begin{abstract} \label{sec:abstract}
The ability to prepare sizeable multi-qubit entangled states with full qubit control is a critical milestone for physical platforms upon which quantum computers are built. We investigate the extent to which entanglement is found within a prepared graph state on the~20-qubit superconducting quantum computer~\textit{IBM Q} \textit{Poughkeepsie}. We prepared a graph state along a path consisting of all twenty qubits within the device and performed full quantum state tomography on all groups of four connected qubits along this path. We determined that each pair of connected qubits was inseparable and hence the prepared state was entangled. Additionally, a genuine multipartite entanglement witness was measured on all qubit subpaths of the graph state and we found genuine multipartite entanglement on chains of up to three qubits. These results represent a demonstration of entanglement in one of the largest solid-state qubit arrays to date and indicate the positive direction of progress towards the goal of implementing complex quantum algorithms relying on such effects.
\end{abstract}

\section{Introduction} \label{sec:introduction}


Quantum entanglement describes non-classical correlations between different subsystems~\cite{horodecki2009quantum}. It is regarded as one of the key hallmarks separating quantum from classical systems. Its significance is fundamental, being the subject of the `spooky' correlations noted by Einstein, Podolsky and Rosen (EPR)~\cite{einstein1935can} and used by Bell~\cite{bell1964einstein, aspect1982experimental, hensen2015loophole} to rule out local hidden variable descriptions of quantum mechanics. More recently the utility of entanglement has become apparent, with quantum entanglement viewed as a useful \emph{resource} to aid information processing tasks~\cite{wootters1998quantum}. The simplest form of entanglement, EPR pairs, enable tasks such as quantum cryptography \cite{ekert1991quantum}, super-dense coding~\cite{bennett1992communication}, teleportation~\cite{bennett1993teleporting} and entanglement swapping~\cite{yurke1992bell, zukowski1993event}. More complex, multi-qubit examples of entanglement enable one-way quantum computation~\cite{raussendorf2001one}, entanglement assisted error correction~\cite{brun2006correcting}, and as a tomographic resource~\cite{nagata2007beating}.

Quantum entanglement often is seen as a key ingredient if quantum computers are to demonstrate an advantage over classical computers. In particular, if a quantum system is not highly entangled it can often be simulated efficiently on a classical computer~\cite{vidal2003efficient, verstraete2004renormalization, verstraete2008matrix}. Multi-qubit entanglement is therefore a fundamental property for potential quantum computers to demonstrate, if they are to ultimately outperform classical computation~\cite{preskill2012quantum}. To this end large multi-qubit entangled states have been demonstrated in a number of experimental systems. On systems with full qubit control, entanglement has been shown on up to a~16-qubit superconducting system~\cite{wang201816} and a~20-qubit ion trap system~\cite{friis2018observation, monz201114}, while genuine multipartite entanglement has been shown on up to an~18-qubit photonic system~\cite{wang201818},~12-photon system~\cite{zhong201812, wang2016experimental} and~12-qubit superconducting system~\cite{PhysRevLett.122.110501, song201710}. Genuine multipartite entanglement has also been shown on arrays of up to 22 atomic ensembles however they are not convenient for the realisation of quantum computation~\cite{pu2018experimental}.

Over the past few years, a series of quantum devices have been released by IBM that comprise five to twenty superconducting qubits and can be accessed via their cloud service~\cite{ibmpress2019ibm, ibmq}. Of particular interest here is the device \textit{IBM Q} \textit{Poughkeepsie}, which exhibits improved error rates over previous devices. In this paper we consider entanglement generation and quantification on the \textit{Poughkeepsie} device. We show that \textit{Poughkeepsie} can be fully entangled. To do this, the system was prepared into a graph state along a path through all twenty qubits and full quantum state tomography performed on all groups of four connected qubits along this path. Using these measurements, we evaluated the negativity~\cite{plenio2005introduction, christandl2006structure} to detect entanglement between each pair of qubits along the chain. By showing that \emph{every} pair is entangled we conclude that the graph state is not separable and hence fully entangled. The magnitude of the entanglement measured between each pair of qubits in the graph state, and the number of qubits entangled, surpasses the entanglement previously measured between pairs within the~16-qubit~\textit{IBM Q Ruschlikon (ibmqx5)} device~\cite{wang201816}. In addition, we consider \emph{genuine multipartite entanglement} within the 20-qubit graph state. To quantify this, we make use of an entanglement witness~\cite{toth2005entanglement}, and show there is genuine multipartite entanglement on chains of three qubits. During the submission process, we were made aware of recently surfaced works that demonstrate genuine multipartite entanglement on~18 and 20-qubit Greenberger-Horne-Zeilinger (GHZ) states~\cite{song2019observation, omran2019generation}, and in particular on an~18-qubit GHZ state prepared on the \textit{IBM Q System One} device~\cite{wei2019verifying}.



\section{Entanglement Measures}

\begin{figure}
     \centering
     \begin{subfigure}[a]{0.48\textwidth}
         \centering
         \subcaption[short for lof]{Separable -- No Entanglement}
         \vspace{0.1cm}
         \includegraphics[width=0.9\textwidth]{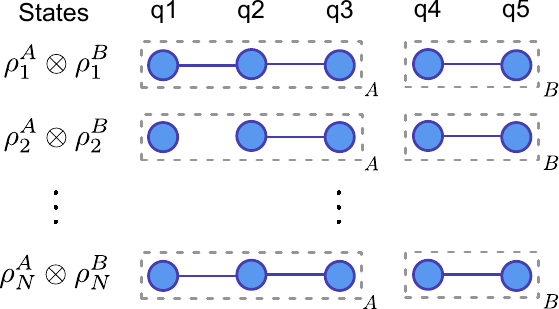}
         \label{fig:separable}
     \end{subfigure}
    \hfill
     \begin{subfigure}[a]{0.48\textwidth}
         \centering
         \subcaption[short for lof]{Biseparable}
         \vspace{0.1cm}
         \includegraphics[width=0.9\textwidth]{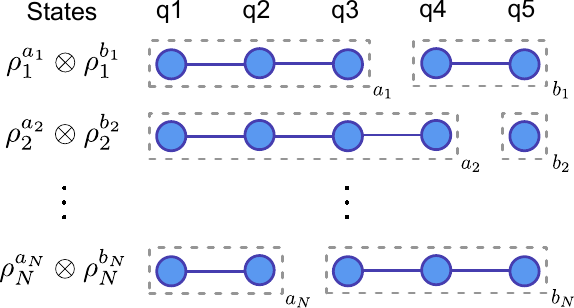}
         \label{fig:bipartite-entanglement}
     \end{subfigure}

     \begin{subfigure}[a]{0.48\textwidth}
     \vspace{0.8 cm}
         \centering
         \subcaption[short for lof]{Not Biseparable}
         \vspace{0.1cm}
         \includegraphics[width=0.9\textwidth]{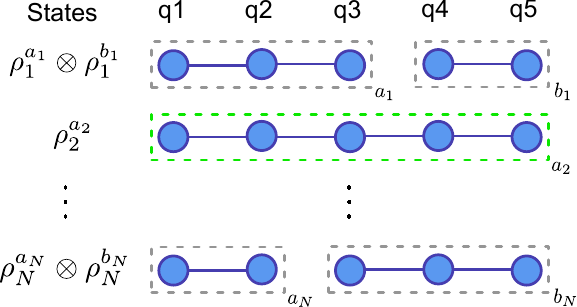}
         \vspace{0.3cm}
         \label{fig:genuine-multipartite-entanglement}
     \end{subfigure}
        \caption{This figure depicts multi-qubit entanglement concepts related to our work. A system is entangled if it is not separable. A system is genuinely multipartite entangled if it is entangled but not biseparable. Vertices represent qubits and edges represent two-qubit entanglement. $N$ is the number of pure states over all qubits in a composition of a mixed state $\rho$. \textbf{(a)} A mixed state $\rho$ is separable if and only if there exists a qubit bipartition $A$ and $B$ such that $\rho = \sum_i p_i \rho_i^{A} \otimes \rho_i^{B}$.~\textbf{(b)} A mixed state~$\rho$ is biseparable if and only if for each state~$i$ there exists bipartitions~$a_i$ and~$b_i$ such that $\rho = \sum_i p_i \rho_i^{a_i} \otimes \rho_i^{b_i}$. \textbf{(c)} An example of a system that is not biseparable.}\label{fig:types-of-entanglement}
\end{figure}

A pure state $\rho_\text{p}$ is \textit{separable} if there exists a qubit bipartition~$A$ and~$B$ such that \begin{equation}
    \rho_\text{p} = \rho^A \otimes \rho^B,
\end{equation}
where $\rho^A$ and $\rho^B$ are pure states of~$A$ and~$B$ respectively. A pure state is \textit{entangled} if it is not separable. Similarly, a mixed state $\rho$ over $n$ qubits is separable (see Fig.~\ref{fig:separable}) if it can be expressed as a probabilistic mixture of separable pure states with respect to a fixed qubit bipartition $A$ and $B$. That is,
\begin{equation}\label{eq:bipartition}
    \rho = \sum \limits_i^N p_i \rho_i^{A} \otimes \rho_i^{B},
\end{equation}
where $\rho^A_i$ and $\rho^B_i$ are pure states of~$A$ and~$B$, $N$ is the number of pure states over all qubits in the composition and the probabilities satisfy~$p_i \geq 0$ and~$\sum_i p_i = 1$. A mixed state is entangled if it is not separable.

Entanglement between bipartitions $A$ and $B$ can be determined by calculating an entanglement measure \cite{vedral1997quantifying, vidal2000entanglement}, such as the negativity, of the state. For a given state $\rho$, the negativity $\mathcal{N}(\rho)$ is given by
\begin{equation}
    \mathcal{N}(\rho) := \sum\limits_i \frac{|\lambda_i| - \lambda_i}{2},
\end{equation}
where $\lambda_i$ are the eigenvalues of the partial transpose of~$\rho$ with respect to the system $B$~\cite{vidal2002computable}. If the negativity is greater than zero, then the two partitions are entangled. In the case of a 2-qubit mixed state, the negativity is zero if and only if the two qubits are separable~\cite{horedecki1996separability}. So although other entanglement measures exist such as concurrence \cite{wootters1998entanglement}, for a 2-qubit system, the negativity alone is a sufficient condition.

For a state with more than two qubits, things are more complicated because each pure state can be separated into different partitions. A mixed state $\rho$ is biseparable (see Fig.~\ref{fig:bipartite-entanglement}) if it can be expressed as
\begin{equation}
    \rho = \sum\limits_i^N p_i \rho_i^{a_i} \otimes \rho_i^{b_i},
\end{equation}
where $N$ is the number of pure states over all qubits in the composition,~$\rho^{a_i}_i$ and~$\rho^{b_i}_i$ are pure states of qubit bipartitions~$a_i$ and~$b_i$, and probabilities satisfy~$p_i \geq 0$ and~$\sum_i p_i = 1$. A system is genuinely multipartite entangled if it is entangled but not biseparable as shown in Fig.~\ref{fig:genuine-multipartite-entanglement}.

Detecting genuine multipartite entanglement using full quantum tomography can be a prohibitively expensive process. However, an \textit{entanglement witness} \cite{terhal2000bell}, which is an observable that detects the presence of entanglement, can require far fewer measurements. An entanglement witness has a non-negative expectation value for all separable states and a negative value for some entangled states. A negative expectation value implies the presence of entanglement, however the converse is not necessarily true. Here, we use a genuine multipartite entanglement witness \cite{toth2005entanglement} which is defined on a graph state $|G_n\rangle$ of $n$ qubits as 
\begin{equation}\label{eq:witness}
    \mathcal{W}^{(G_n)} := (n-1)\mathbb{1} - \sum \limits_i S_i^{(G_n)}
\end{equation}
where $S_i^{(G_n)}$ is the $i^{\text{th}}$ generator of the stabiliser group of $|G_n\rangle$.

To show that all qubits within a quantum computer can be entangled, we first aim to prepare them into a highly entangled state. We use the graph state since it has been shown to have entanglement that is more robust against local measurements and noise than GHZ states \cite{briegel2001persistent}. A graph state is defined in relation to a graph where each qubit corresponds to a vertex and is prepared in the state $|+\rangle \equiv (|0\rangle + |1\rangle)/\sqrt{2}$, then controlled-phase gates are applied between adjacent qubits within the graph. The state is genuinely multipartite entangled in the absence of errors and decoherence. The resulting state has the form
    \begin{equation}
        |G_n\rangle = \prod \limits_{(\alpha,\beta)\in E} \mathrm{CZ}^\alpha_{\beta} \;|+\rangle^{\otimes n},
    \end{equation}
where $E$ is the edge set of the graph $G_n$ corresponding to the $n$ qubit graph state, and $\mathrm{CZ}^\alpha_{\beta}$ represents a controlled-phase gate between adjacent qubits~$\alpha$ and~$\beta$. The set of generators of the stabiliser group, which we refer to as stabilisers, for the graph state can be written as
    \begin{equation}
        K_\alpha = X^\alpha \prod_{\beta \in N_\alpha} Z^\beta,
    \end{equation}
where $N_\alpha$ is the set of neighbouring qubits of $\alpha$.

\begin{figure}
    \centering
    \includegraphics[scale=0.8]{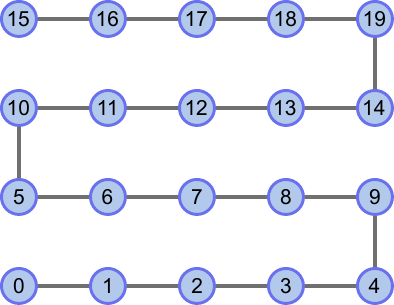}
    \caption{Embedded graph state within the \textit{IBM Q} \textit{Poughkeepsie} layout. Nodes represent qubits and edges represent two-qubit gates used to prepare the graph state. Actual layout connectivity is not shown.}\label{fig:poughkeepsie-device}
\end{figure}

\begin{figure}
     \centering
     
     \begin{subfigure}[a]{0.48\textwidth}
         \centering
         \subcaption[short for lof]{}
         \includegraphics[height=320 pt]{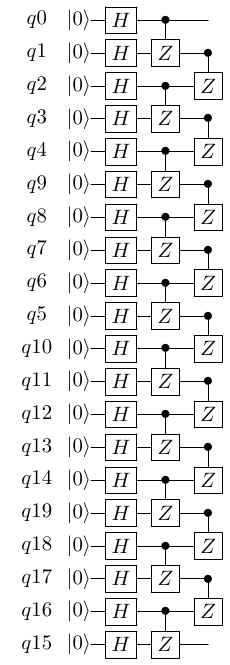}
     \end{subfigure}
     \hfill
     \begin{subfigure}[a]{0.48\textwidth}
         \centering
         \subcaption[short for lof]{}
         \includegraphics[height=320 pt]{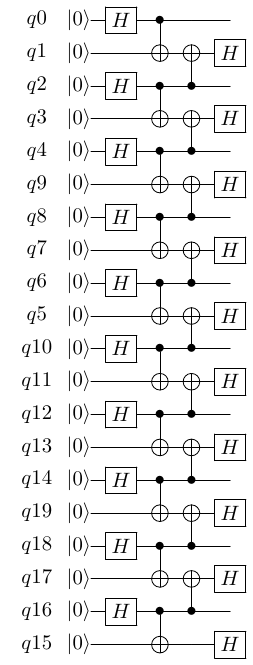}
     \end{subfigure}
     
        \caption{\textbf{(a)} Circuit for the 20-qubit graph state. \textbf{(b)} The same circuit is reduced to CNOTs so that it can be directly performed on the quantum device without further optimisation.}\label{fig:graph-state-circuit}
\end{figure}

Graph states have the property that by projecting all but 2 qubits to the $Z$-basis we are in principle left with maximally entangled Bell states (up to local transformations). The entanglement between these remaining qubits may then be determined by measuring the negativity.

As an example, consider a 4-qubit graph state on qubits $1, 2, 3$ and $4$ which is stabilised by the operators
    \begin{align}
        K_1 &= X_1 Z_2 I_3 I_4, \nonumber \\
        K_2 &= Z_1 X_2 Z_3 I_4, \nonumber\\
        K_3 &= I_1 Z_2 X_3 Z_4, \text{ and } \nonumber\\
        K_4 &= I_1 I_2 Z_3 X_4.
    \end{align}
Since $K_1$ anti-commutes with $Z_1 I_2 I_3 I_4$ while the other stabilisers commute, projecting qubit $1$ onto the $Z$-basis corresponds to replacing $K_1$ with $K_1^\prime = (-1)^{m_1} Z_1 I_2 I_3 I_4$ \cite{nielsen2002quantum} where $m_i \in \{0, 1\}$ is the projected state of qubit~$i$. Similarly, projecting qubit $4$ onto the $Z$-basis corresponds to replacing $K_4$ with $K_4^\prime = (-1)^{m_4} I_1 I_2 I_3 Z_4$. The stabilisers can then be simplified to
\begin{align}
    K_\beta &= (-1)^{m_1} X_2 Z_3, \text{ and }\nonumber\\
    K_\gamma &= (-1)^{m_4} Z_2 X_3,
\end{align}
which stabilise the Bell states (up to local transformations)
\begin{align}
    \frac{1}{\sqrt{2}}\Big[|0+\rangle_{2,3} \pm |1-\rangle_{2,3}\Big],\nonumber\\
    \frac{1}{\sqrt{2}}\Big[|1+\rangle_{2,3} \pm |0-\rangle_{2,3}\Big].
\end{align}

Since the dimension of the Hilbert space doubles for each qubit, performing full quantum state tomography on a 20-qubit system would require approximately~3.5 billion measurements. However, to show that a graph state is entangled, we can show that for any given bipartition of the graph, qubits in one partition are entangled with those in the other. For a connected graph, any bipartition will have at least one qubit in one partition that has a neighbour in the other partition. Thus, if we can show that there is entanglement between every pair of connected qubits, then any two partitions must contain a qubit in one partition that is entangled with a qubit from the other partition. Using local operations, we project the pair into a Bell state, and then perform quantum tomography on each pair of connected qubits $\alpha$ and $\beta$ with their neighbours $N_\alpha$ and $N_{\beta}$, then calculate the negativity between them to establish entanglement between each pair.

\subsection{20-qubit Entanglement}

\begin{figure}
     \centering
     \begin{subfigure}[a]{0.48\textwidth}
         \centering
         \subcaption[short for lof]{Negativity from Zero State Projection}\label{fig:zero-state-negativity}
         \includegraphics[width=1\textwidth]{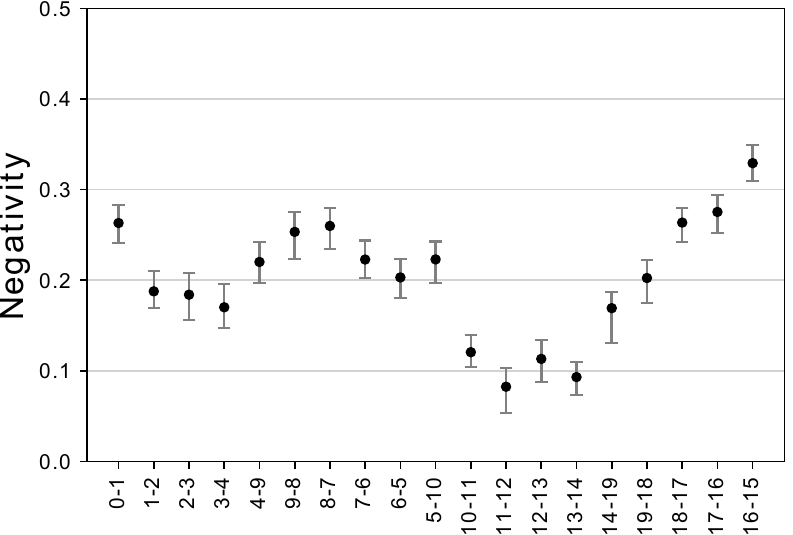}
     \end{subfigure}
     \hfill
     \begin{subfigure}[a]{0.48\textwidth}
         \centering
         \subcaption[short for lof]{Largest Negativity over Projections}\label{fig:largest-negativity}
         \includegraphics[width=1\textwidth]{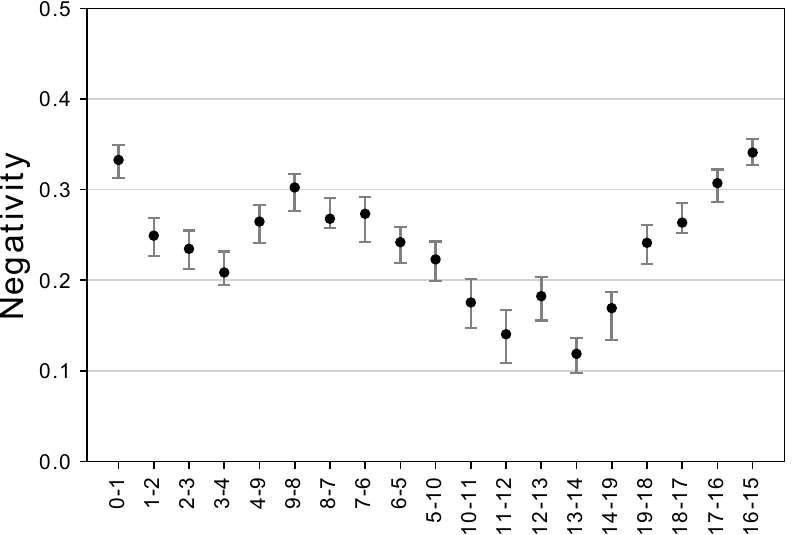}
     \end{subfigure}
     \hfill
     \vspace{5pt}
     \begin{subfigure}[a]{0.48\textwidth}
         \centering
         \subcaption[short for lof]{Mean Negativity over Projections}\label{fig:mean-negativity}
         \includegraphics[width=1\textwidth]{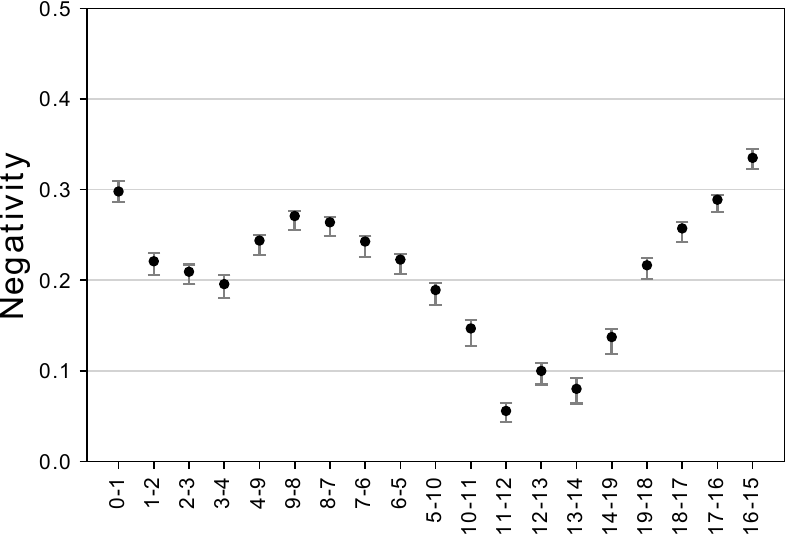}
     \end{subfigure}
        \caption{Graph state negativity values between connected pairs~$i$-$j$ of qubits~$i$ and~$j$. Negativity ranges between~0 and~0.5 where larger values represent more entanglement. The~95$\%$ confidence intervals are estimated using bootstrapping methods. \textbf{(a)} The neighbouring qubits are projected onto the zero state of the Z-basis which acts as a point of comparison for previous work on the 16-qubit~\textit{IBM Q Ruschlikon (ibmqx5)} device~\cite{wang201816}. \textbf{(b)} The largest negativity of all combinations of neighbour projections onto the Z-basis. \textbf{(c)} The average negativity of all combinations of neighbour projections onto the Z-basis.}\label{fig:negativity}
\end{figure}

\begin{figure*}
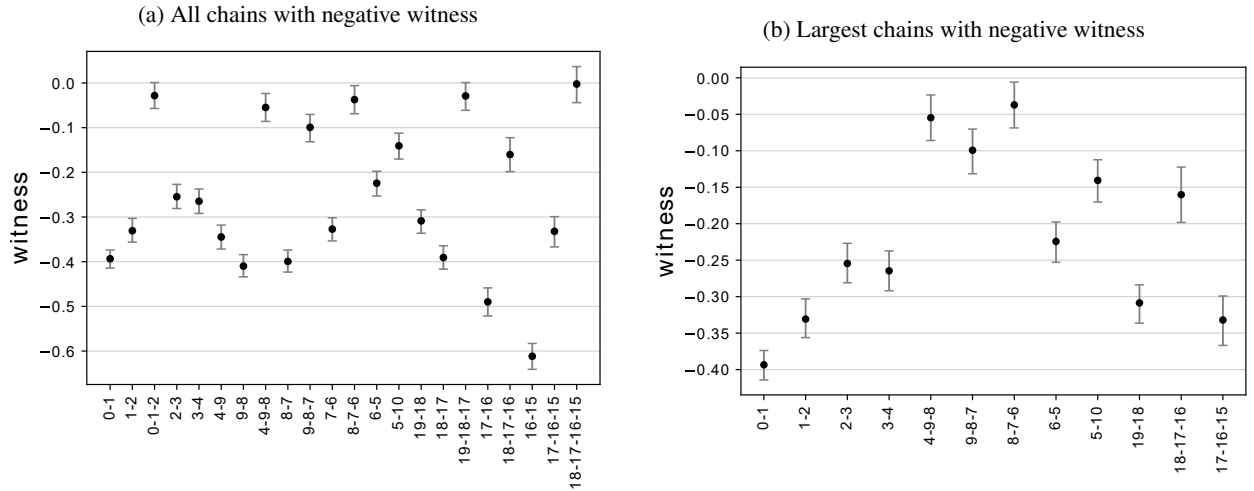

\centering
        {\rowcolors{2}{white}{gray!15}
            \begin{tabular}{ c|c|c|c }
                Qubit pairs & Zero state projection & Largest over projections & Mean over projections\\ [0.5ex]
                \hline
                 15-16 & 0.329 (0.310, 0.349) & 0.341 (0.327, 0.355) & 0.335 (0.322, 0.345) \\ 
                 16-17 & 0.275 (0.252, 0.294) & 0.307 (0.286, 0.322) & 0.289 (0.275, 0.294)\\  
                 17-18 & 0.264 (0.243, 0.280) & 0.264 (0.252, 0.285) & 0.257 (0.242, 0.264)\\
                 18-19 & 0.202 (0.175, 0.222) & 0.241 (0.218, 0.261) & 0.216 (0.202, 0.224)\\
                 19-14 & 0.169 (0.131, 0.187) & 0.169 (0.134, 0.187) & 0.137 (0.119, 0.146)\\
                 14-13 & 0.093 (0.073, 0.110) & 0.119 (0.097, 0.136) & 0.080 (0.064, 0.092)\\
                 13-12 & 0.113 (0.087, 0.134) & 0.182 (0.156, 0.203) & 0.100 (0.085, 0.108)\\
                 12-11 & 0.082 (0.054, 0.103) & 0.140 (0.109, 0.167) & 0.056 (0.044, 0.065)\\
                 11-10 & 0.121 (0.104, 0.140) & 0.175 (0.147, 0.202) & 0.147 (0.128, 0.156)\\
                 10-5 & 0.223 (0.197, 0.243) & 0.223 (0.199, 0.243) & 0.189 (0.172, 0.197)\\
                 5-6 & 0.203 (0.181, 0.223) & 0.242 (0.219, 0.258) & 0.223 (0.207, 0.229)\\
                 6-7 & 0.223 (0.202, 0.244) & 0.273 (0.242, 0.291) & 0.243 (0.226, 0.249)\\
                 7-8 & 0.260 (0.235, 0.280) & 0.268 (0.258, 0.290) & 0.264 (0.249, 0.270)\\
                 8-9 & 0.253 (0.223, 0.275) & 0.302 (0.277, 0.317) & 0.271 (0.255, 0.276)\\
                 9-4 & 0.220 (0.197, 0.242) & 0.265 (0.241, 0.283) & 0.244 (0.228, 0.250)\\
                 4-3 & 0.170 (0.148, 0.196) & 0.208 (0.195, 0.232) & 0.196 (0.180, 0.205)\\
                 3-2 & 0.184 (0.156, 0.208) & 0.235 (0.213, 0.255) & 0.209 (0.196, 0.217)\\
                 2-1 & 0.188 (0.170, 0.210) & 0.249 (0.227, 0.268) & 0.221 (0.206, 0.230)\\
                 1-0 & 0.263 (0.241, 0.283) & 0.333 (0.313, 0.349) & 0.298 (0.286, 0.309)\\
            \end{tabular}
        }
        \caption{Graph state negativity between connected pairs $i$-$j$ of qubits $i$ and $j$ and 95\% confidence intervals estimated using bootstrapping methods. The zero state projection column refers to negativity values for qubit pairs with their neighbours projected onto the zero state of the $Z$-basis (shown in Fig.~\ref{fig:zero-state-negativity}). The largest over projections column refers to the largest negativity over projecting neighbours onto each state of the $Z$-basis (shown in Fig.~\ref{fig:largest-negativity}). The mean over projections column refers to the mean negativity over projecting neighbours onto each state of the $Z$-basis (shown in Fig.~\ref{fig:mean-negativity}).}\label{fig:negativity-table}
\end{figure*}

The strategy to detect entanglement along a chain of qubits was implemented on physical hardware, specifically that of \textit{IBM Q} \textit{Poughkeepsie}. The device consists of 20 superconducting qubits that are capable of coherence times of $T_1$, $T_2$ $\sim$ 100 $\mu$s \cite{ibmblog2019cramming}. A path consisting of all qubits was embedded onto the device layout as shown in Fig~\ref{fig:poughkeepsie-device} and a corresponding graph state was prepared using the circuit shown in Fig.~\ref{fig:graph-state-circuit}.

To determine that each pair of connected qubits are entangled, we perform full quantum state tomography on each quad~--~the pair and their neighbours -- as described in the previous section. When constructing the tomography circuits, measurements that include the $I$ basis can be post-processed, totalling just~$3^n$ circuit configurations instead of the full~$4^n$ configurations. A total of~2048 shots were used for each measurement. To ensure that the constructed density matrix has physical (non-negative) eigenvalues, the nearest physical state under the 2-norm is determined using the efficient procedure introduced by Smolin \textit{et al.}~\cite{smolin2012efficient}. 

The negativity for each~$Z$-basis state projection was calculated and the results are plotted in Fig.~\ref{fig:negativity} and tabulated in Table~\ref{fig:negativity-table}. The negativity ranges between~0 and~0.5 where~0 indicates no entanglement and~0.5 indicates maximum entanglement. There are four different, equally likely, outcomes for Z-measurements on the neighbouring qubits in the chain (and two different outcomes for the pairs at the end of each chain). Each of these measurement outcomes leads to a different Bell state, which, being noisy, can have a different negativity. Fig.~\ref{fig:negativity} shows for every pair of neighbouring qubits in the chain (a) the negativity of the~00 projection/measurement result, (b) the maximum of the four negativities and (c) the mean of the four negativities. Our results, in all three cases, indicate that the \textit{Poughkeepsie} device has been fully entangled.

The negativities for the zero state projection shown in Fig.~\ref{fig:zero-state-negativity} were used to compare entanglement with the previous results found within the~16-qubit \textit{IBM Q Ruschlikon (ibmqx5)} device \cite{wang201816}. Rather than embedding a path, they embed a loop which has the benefit that instead of requiring all connected pairs to have statistically significant non-zero negativity, it allows up to one pair to have a non-significant value. For the 16-qubit graph state, 15 of the 16 connected pairs had statistically significant non-zero negativity, so we compare with those. We found that the magnitude of entanglement between pairs of qubits in the \textit{IBM Q Poughkeepsie} device surpasses the \textit{Ruschlikon} device. We found that the minimum and maximum negativities and~95\% confidence intervals calculated on the \textit{IBM Q Poughkeepsie} device were~0.082~(0.054, 0.103) and~0.329~(0.310, 0.349) while the \textit{Ruschlikon} device had~0.034~(0.012, 0.053) and~0.241~(0.229, 0.261) respectively. The~95$\%$ confidence intervals are calculated using bootstrapping methods~\cite{efron1986bootstrap}. While the results imply entanglement in the 20-qubit system they do not allow us to rule out the possibility that the state is biseparable.

\begin{figure}
     \centering
     \begin{subfigure}[a]{0.48\textwidth}
         \centering
         \subcaption[short for lof]{All chains with negative witness}\label{fig:witness_ibmq_poughkeepsie-all}
         \includegraphics[width=1\textwidth]{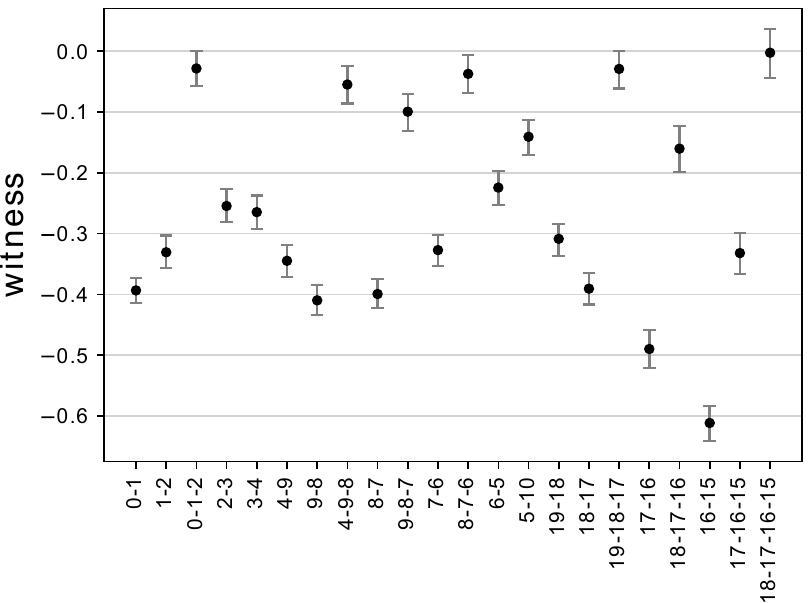}
     \end{subfigure}
     \hfill
     \begin{subfigure}[a]{0.48\textwidth}
         \centering
         \subcaption[short for lof]{Largest chains with negative witness}\label{fig:witness_ibmq_poughkeepsie-reduced}
         \includegraphics[width=1\textwidth]{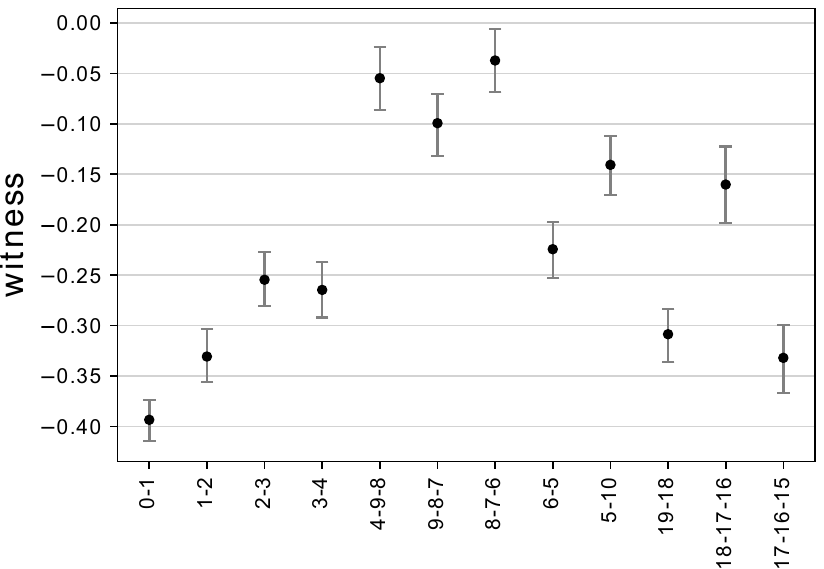}
     \end{subfigure}
        \caption{Genuine multipartite entanglement witness expectation values calculated on the full 20-qubit graph state, where $i$-$...$-$j$ denotes the qubit chain between qubits $i$ and $j$. Negative values imply genuine multipartite entanglement. Values that are above zero have been omitted for readability. Up to~3-qubit chains were found to be genuinely multipartite entangled and are visually summarised in Fig.~\ref{fig:poughkeepsie-device-gme}. A~4-qubit chain between qubits 15 and 18 was found to have a negative value but was non-significant. The 95$\%$ confidence intervals are estimated using bootstrapping methods. \textbf{(a)} Values that are above zero have been omitted for readability. \textbf{(b)} Values that are non-significant or redundant have been omitted, e.g. if $8$-$7$-$6$ are genuinely multipartite entangled then $8$-$7$ must also be genuinely multipartite entangled making the result for $8$-$7$ redundant.}\label{fig:witness_ibmq_poughkeepsie}
\end{figure}

\begin{figure}
    \centering
    \includegraphics[scale=0.8]{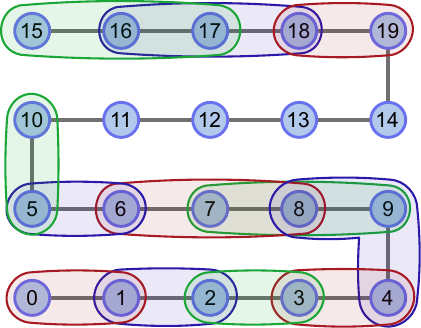}
    \caption{Regions of genuine multipartite entanglement found within the 20-qubit graph state using an entanglement witness. There were 5 chains of 3 qubits that were found to be genuinely multipartite entangled. The witness expectation values are shown in Fig.~\ref{fig:witness_ibmq_poughkeepsie}. The colours are only used to help visually distinguish group.}\label{fig:poughkeepsie-device-gme}
\end{figure}

\subsection{Genuine Multipartite Entanglement}
To further investigate the entanglement and biseparability of the graph state, a genuine multipartite entanglement witness is measured. Using the tomography results obtained for the~20-qubit graph state, the genuine multipartite entanglement witness in Eq.~\ref{eq:witness} was calculated for all chains of qubits along the graph state. The results are shown in Fig.~\ref{fig:witness_ibmq_poughkeepsie}, where negative values imply genuine multipartite entanglement and the 95$\%$ confidence intervals are estimated using bootstrapping methods~\cite{efron1986bootstrap}. Up to~3-qubit chains were found to be genuinely multipartite entangled and are visually summarised in Fig.~\ref{fig:poughkeepsie-device-gme}. A~4-qubit chain between qubits~15 and~18 was found to have a negative witness however was non-significant. Figure~\ref{fig:witness_ibmq_poughkeepsie-all} shows all values below zero, while Fig~\ref{fig:witness_ibmq_poughkeepsie-reduced} has non-significant and redundant values omitted for clarity, e.g. if~$(8$-$6)$ are genuinely multipartite entangled then~$(8$-$7)$ must also be genuinely multipartite entangled making the result for~$(8$-$7)$ redundant. The witness found no entanglement for all pairs of qubits between qubit~10 and~19 even though they were shown to be entangled by the negativity analysis. This is due to the witness detecting genuine multipartite entanglement in only some genuinely multipartite entangled states. However in the case of negativity, for all~2-qubit states, non-zero negativity is equivalent to entanglement. These results are inconclusive as to whether the graph state is genuinely multipartite entangled. To investigate further, a more tailored experiment could be used such as calculating the fidelity~\cite{guhne2009entanglement} or entanglement structure~\cite{lu2018entanglement} of a GHZ state.

\section{Conclusion}
By preparing a graph state along a path consisting of all 20 qubits within the \textit{IBM Q} \textit{Poughkeepsie} device, we were able to determine that entanglement is present between any bipartition of the system in this state. To do this we performed full quantum state tomography over each connected pair and neighbours. By calculating the negativity between every pair, we determined that each pair possessed non-zero entanglement. We found that the level of entanglement of the full system surpasses previous results found in the~16-qubit \textit{IBM Q Ruschlikon (ibmqx5)} device~\cite{wang201816}. Additionally, we calculated a genuine multipartite entanglement witness along all qubit sub-paths of the~20-qubit graph state, finding genuine multipartite entanglement in up to~3-qubit chains. These results indicate that the ability to entangle qubits in these devices is steadily improving as required for the physical implementation of complex quantum algorithms.
\section*{Acknowledgements}
This work was supported by the University of Melbourne through the establishment of an IBM Network Q Hub at the University. We would like to thank Anna Phan and Frank Suits for reading through this paper and providing valuable feedback.




%
%

\bibliographystyle{unsrt}

\end{document}